\documentclass[11pt]{article}

\usepackage[margin=1in]{geometry}
\usepackage{amsmath,amssymb,amsthm}
\usepackage{booktabs}
\usepackage{array}
\usepackage{microtype}
\usepackage[dvipsnames]{xcolor}
\usepackage[colorlinks=true,linkcolor=MidnightBlue,citecolor=MidnightBlue,urlcolor=MidnightBlue]{hyperref}
\usepackage{enumitem}
\usepackage{setspace}

\newtheorem{proposition}{Proposition}

\newcommand{\E}{\mathbb{E}}
\newcommand{\Var}{\operatorname{Var}}

\title{Fertility under Career Risk:\\Limited Commitment and Insurance against the Child Penalty}
\date{}
\author{Ruiwu LIU}

\begin{document}
\maketitle

\begin{abstract}
Couples decide whether to have children before knowing the persistent career loss generated by parenthood.  After the loss is realized, household resources are renegotiated and individual market earnings partly determine outside options.  This limited-commitment friction places a disproportionate share of female earnings risk on the wife even when expected household resources are unchanged.  We embed this mechanism in a generalized-Nash fertility bargain.  The model has a unique positive agreement whenever both spouses have positive marginal gains at zero fertility.  A larger expected child penalty and greater career-loss risk reduce agreed fertility.  A higher loading of own earnings into outside options reduces fertility precisely when the wife's participation margin dominates the husband's corresponding gain.  Insurance against the unexpected component of the career loss raises fertility without subsidizing its mean, whereas a conventional per-child transfer operates only through expected cost.  A full-commitment benchmark isolates the risk-sharing distortion.  In an illustrative discrete-parity exercise, rising career risk lowers desired family size from three children to two and then one, while partial earnings-loss insurance can restore the three-child outcome.  These results identify a policy margin that is not available in a deterministic participation-constraint model: insurance against the distribution of the child penalty.
\end{abstract}

\noindent\textbf{Draft status.} This document completes the analytical core and verifies its mechanisms numerically.  The numerical exercise is illustrative, not a calibration.  Empirical discipline and a quantitatively specified policy environment are the next required steps before journal submission.

\section{Motivation and contribution}

Most household fertility models summarize the child penalty by its expected level.  That is natural when fertility and transfers are chosen under complete information.  It misses a different margin: a couple may decide on fertility before learning how severely parenthood will affect the mother's persistent earnings, promotion path, occupational attachment, or future employability.  The household then has to allocate an ex post loss that could not be contracted on perfectly ex ante.

This paper studies that uncertainty.  Three ingredients are sufficient.

\begin{enumerate}[label=(\roman*)]
    \item Fertility is agreed before a persistent female career loss is known.
    \item Consumption is renegotiated after the loss is realized.
    \item Own market earnings affect outside options, so the realized loss changes both total resources and the intrahousehold allocation.
\end{enumerate}

The mechanism differs from the companion static paper.  There, a known participation constraint determines the transfer required to reach a fertility agreement.  Here, the missing market is state-contingent insurance: the couple cannot fully commit before birth to compensate the spouse who later bears the career shock.  The main state variable is therefore the \emph{distribution} of the child penalty, not only its expected value.

The distinction also separates the proposed paper from recent dynamic bargaining models in which a deterministic or estimated child penalty changes outside options and bargaining power.  Examples include Doepke and Kindermann's negotiation model, Sakamoto's dynamic collective labor-supply model, and Xie's life-cycle limited-commitment model.  The intended contribution here is narrower and analytically transparent: to identify the uninsurable career-risk wedge, derive its incidence inside the household, and compare insurance against that risk with an ordinary child subsidy.  Aggregate uninsurable earnings risk has appeared in quantitative fertility models, notably Sommer (2016); the present model instead makes the risk fertility-specific and lets limited commitment determine which spouse bears it.

\section{Environment}

\subsection{Timing}

There are two spouses, a husband $m$ and a wife $w$.  Fertility $n\geq 0$ is continuous in the analytical model and integer-valued in the numerical parity extension.  The timing is:

\begin{enumerate}
    \item Before birth, the spouses bargain over $n$.
    \item Conditional on $n$, a persistent career loss per child, $X$, is realized:
    \begin{equation}
        X=\mu+\varepsilon, \qquad \E[\varepsilon]=0,
        \qquad \Var(\varepsilon)=\sigma^2.
        \label{eq:risk}
    \end{equation}
    \item Household consumption is renegotiated after $X$ is observed.  Ex ante agreements about the division of resources are not fully enforceable.
\end{enumerate}

The husband's market income is $a_m$.  The wife's post-birth market income is
\begin{equation}
    y_w(n,X)=a_w-Xn.
\end{equation}
There is a direct resource cost $kn$ and a marital surplus $M$.  Baseline resources are assumed sufficiently large that consumption remains positive in the relevant states.  Because the analysis uses gains relative to childlessness, $a_m$, $a_w$, and $M$ drop out of the reduced-form fertility problem.

\subsection{Ex post renegotiation and loss incidence}

Let $\lambda\in[0,1]$ measure how strongly own market earnings are protected by outside options in ex post renegotiation.  Let $\omega\in(0,1)$ be the wife's share of the remaining marital surplus.  Define
\begin{equation}
    S(n,X)=M+(1-\lambda)\bigl[a_m+y_w(n,X)\bigr]-kn.
\end{equation}
The renegotiated allocation is
\begin{align}
    c_w(n,X)&=\lambda y_w(n,X)+\omega S(n,X),\label{eq:cw}\\
    c_m(n,X)&=\lambda a_m+(1-\omega)S(n,X).\label{eq:cm}
\end{align}
Equations \eqref{eq:cw}--\eqref{eq:cm} exhaust household resources.  More importantly, they imply that the incidence of one unit of the wife's career loss is
\begin{align}
    \theta_w&=\omega+\lambda(1-\omega),\label{eq:thetaw}\\
    \theta_m&=(1-\omega)(1-\lambda),\label{eq:thetam}
\end{align}
with $\theta_w+\theta_m=1$.  Under fixed sharing ($\lambda=0$), the wife bears $\omega$ and the husband bears $1-\omega$.  As own earnings matter more for outside options, an additional share $\lambda(1-\omega)$ is shifted from the husband to the wife.

This is the paper's core incidence result.  Limited commitment does not create the aggregate career loss; it determines who remains exposed to the loss after household renegotiation.

\subsection{Ex ante gains from fertility}

Spouse $i\in\{m,w\}$ receives benefit $b_i\log(1+n)$.  In addition, the wife bears a deterministic nonmarket burden $\delta n$.  Preferences over the random payoff are represented in mean--variance form, with risk-aversion parameter $\eta_i>0$.  The representation is exact as a certainty-equivalent calculation for normally distributed monetary risk under constant absolute risk aversion; here it is used as the cardinal surplus in the bargaining problem.

Relative to childlessness, the spouses' ex ante gains are
\begin{align}
    G_w(n)&=b_w\log(1+n)-C_wn-\frac{R_w}{2}n^2,\label{eq:Gw}\\
    G_m(n)&=b_m\log(1+n)-C_mn-\frac{R_m}{2}n^2,\label{eq:Gm}
\end{align}
where
\begin{align}
    C_w&=\delta+\omega k+\theta_w\mu,
    &R_w&=\eta_w\theta_w^2\sigma^2,\label{eq:coefw}\\
    C_m&=(1-\omega)k+\theta_m\mu,
    &R_m&=\eta_m\theta_m^2\sigma^2.\label{eq:coefm}
\end{align}

The expected loss is linear in its incidence share $\theta_i$, while the risk cost is quadratic in $\theta_i$.  Consequently, concentrating exposure on one spouse can be costly even when the spouses together bear exactly one unit of the aggregate shock.

\section{The fertility bargain}

Let $\rho\in(0,1)$ be the husband's Nash bargaining weight.  Childlessness is the disagreement outcome.  The ex ante agreement solves
\begin{equation}
    n^*\in\arg\max_{n\geq0}\;
    G_m(n)^\rho G_w(n)^{1-\rho}
    \quad\text{subject to}\quad G_m(n)\geq0,\;G_w(n)\geq0.
    \label{eq:nash}
\end{equation}
If no positive fertility level benefits both spouses, $n^*=0$.  Otherwise the solution is interior and satisfies
\begin{equation}
    F(n)\equiv
    \rho\frac{G_m'(n)}{G_m(n)}
    +(1-\rho)\frac{G_w'(n)}{G_w(n)}=0.
    \label{eq:foc}
\end{equation}

\begin{proposition}[Existence and uniqueness]
For each spouse,
\begin{equation}
    G_i''(n)=-\frac{b_i}{(1+n)^2}-R_i<0.
\end{equation}
If $b_i\leq C_i$ for either spouse, no positive $n$ produces a mutually beneficial agreement and $n^*=0$.  If $b_m>C_m$ and $b_w>C_w$, the bargaining problem has a unique positive solution characterized by \eqref{eq:foc}.
\end{proposition}

Strict concavity is useful economically as well as technically.  Multiple mathematical roots are not needed to produce large fertility responses.  With integer parity, a smooth change in risk can move the household discontinuously between adjacent family sizes even though the underlying continuous agreement is unique.

\section{Career losses, risk, and limited commitment}

\begin{proposition}[Mean and risk effects]
Suppose a positive fertility agreement exists.  Then
\begin{equation}
    \frac{\partial n^*}{\partial\mu}<0,
    \qquad
    \frac{\partial n^*}{\partial\sigma^2}<0.
    \label{eq:basiccs}
\end{equation}
A larger expected career loss and a mean-preserving increase in career-loss risk both reduce agreed fertility.
\end{proposition}

The two comparative statics are conceptually distinct.  The first is an expected-resource effect.  The second survives even when expected household income is held fixed.  This second effect is the new Paper B margin.

The effect of $\lambda$ is more informative than an unconditional sign.  Define
\begin{align}
    H_i&\equiv\frac{G_i'}{G_i},\\
    A_i&\equiv\frac{G_i-nG_i'}{G_i^2}>0,\\
    B_i&\equiv\frac{n\left(G_i-\tfrac12nG_i'\right)}{G_i^2}>0,\\
    D_i&\equiv\mu A_i+2\eta_i\theta_i\sigma^2B_i>0,
    \label{eq:Di}
\end{align}
all evaluated at $n^*$.  The quantities $A_i$ and $B_i$ measure how sensitive spouse $i$'s normalized marginal surplus is to expected cost and risk, respectively.

\begin{proposition}[Limited-commitment incidence]
At an interior agreement,
\begin{equation}
    \operatorname{sign}\left(\frac{\partial n^*}{\partial\lambda}\right)
    =\operatorname{sign}\left[\rho D_m-(1-\rho)D_w\right].
    \label{eq:lambda-sign}
\end{equation}
Therefore a stronger loading of own earnings into outside options lowers fertility if and only if
\begin{equation}
    (1-\rho)D_w>\rho D_m.
    \label{eq:wife-margin}
\end{equation}
\end{proposition}

Condition \eqref{eq:wife-margin} has a direct interpretation.  Increasing $\lambda$ moves both the expected career loss and its risk from the husband to the wife.  Fertility falls when the wife's participation margin is more sensitive than the husband's offsetting gain.  This is likely when the wife is more risk averse, already bears a larger nonmarket burden, receives lower fertility benefit, or is closer to her participation boundary.  The model thus predicts heterogeneity rather than imposing that every increase in female outside-option sensitivity must reduce fertility.

\section{Insurance versus a child subsidy}

Consider an actuarially fair contract that insures a fraction $q\in[0,1]$ of the unexpected career innovation $\varepsilon$.  The residual innovation is $(1-q)\varepsilon$, so
\begin{equation}
    R_i(q)=\eta_i\theta_i^2(1-q)^2\sigma^2.
    \label{eq:insurance}
\end{equation}
This policy is a stylized representation of earnings-loss insurance, a return-to-career guarantee, or a state-contingent replacement payment.  It is deliberately defined not to subsidize the expected loss $\mu$.

For comparison, let $s$ be a conventional per-child transfer that reduces the direct expected cost to $k-s$.  It changes
\begin{equation}
    C_w(s)=\delta+\omega(k-s)+\theta_w\mu,
    \qquad
    C_m(s)=(1-\omega)(k-s)+\theta_m\mu,
    \label{eq:cash}
\end{equation}
but leaves $R_i$ unchanged.

\begin{proposition}[Policy comparative statics]
At an interior agreement and for $q<1$,
\begin{equation}
    \frac{\partial n^*}{\partial q}>0,
    \qquad
    \frac{\partial n^*}{\partial s}>0.
    \label{eq:policycs}
\end{equation}
Insurance raises fertility by compressing the distribution of the career penalty.  A child subsidy raises fertility by lowering its expected resource cost.  The two instruments are not equivalent unless risk is absent or households are risk neutral.
\end{proposition}

Proposition 4 does not by itself rank the policies per dollar of public expenditure.  Such a ranking requires a financing rule, take-up, moral hazard, administrative cost, and a calibrated shock distribution.  Its value is to establish that a policy can affect fertility even when it leaves the mean child penalty unchanged.

\section{Full-commitment benchmark}

Suppose the spouses can commit before birth to state-contingent transfers after $X$ is observed.  Let $\vartheta_w$ and $\vartheta_m$ be the shares of the aggregate career innovation borne by the two spouses, with $\vartheta_w+\vartheta_m=1$.  Efficient risk sharing minimizes
\begin{equation}
    \eta_w\vartheta_w^2+\eta_m\vartheta_m^2.
\end{equation}
The solution is
\begin{equation}
    \vartheta_w^E=\frac{\eta_m}{\eta_w+\eta_m},
    \qquad
    \vartheta_m^E=\frac{\eta_w}{\eta_w+\eta_m},
    \label{eq:efficientshares}
\end{equation}
and the aggregate risk coefficient is
\begin{equation}
    R_E=\frac{\eta_w\eta_m}{\eta_w+\eta_m}\sigma^2.
    \label{eq:RE}
\end{equation}

With utilitarian weights normalized to one, the full-commitment fertility surplus is
\begin{equation}
    G_E(n)=B\log(1+n)-C_En-\frac{R_E}{2}n^2,
    \quad B=b_m+b_w,
    \quad C_E=\delta+k+\mu.
    \label{eq:GE}
\end{equation}
If $B>C_E$, efficient fertility solves
\begin{equation}
    \frac{B}{1+n_E}=C_E+R_En_E.
\end{equation}
For $R_E>0$,
\begin{equation}
    n_E=
    \frac{-(C_E+R_E)+
    \sqrt{(C_E+R_E)^2+4R_E(B-C_E)}}{2R_E}.
    \label{eq:nE}
\end{equation}
When $R_E=0$, $n_E=B/C_E-1$.

The benchmark isolates two distortions in the decentralized agreement.  First, ex post renegotiation can concentrate the career shock on the spouse least able to absorb it.  Second, the ex ante fertility bargain is restricted by both individual participation gains.  The comparison is not a claim that all limited-commitment allocations produce inefficiently low fertility for every parameterization; it is a transparent way to measure the wedge in a quantitative application.

\section{Numerical prototype}

\subsection{Baseline}

Table \ref{tab:parameters} reports a normalized illustrative parameterization.  It is selected to demonstrate the mechanisms, not to match a country or cohort.

\begin{table}[htbp]
\centering
\caption{Illustrative baseline parameters}
\label{tab:parameters}
\begin{tabular}{lcl@{\qquad}lcl}
\toprule
Parameter & Value & Interpretation & Parameter & Value & Interpretation\\
\midrule
$b_m$ & 0.65 & husband's child benefit & $b_w$ & 0.50 & wife's child benefit\\
$\delta$ & 0.10 & wife's nonmarket burden & $k$ & 0.15 & direct cost\\
$\mu$ & 0.25 & mean career loss & $\sigma$ & 0.20 & career-risk s.d.\\
$\omega$ & 0.50 & wife's surplus share & $\lambda$ & 0.60 & outside-option loading\\
$\eta_m$ & 2.00 & husband's risk aversion & $\eta_w$ & 3.00 & wife's risk aversion\\
$\rho$ & 0.50 & husband's Nash weight & & &\\
\bottomrule
\end{tabular}
\end{table}

The implied loss-incidence shares are $\theta_m=0.20$ and $\theta_w=0.80$.  Efficient state-contingent sharing would instead give $\vartheta_m^E=0.60$ and $\vartheta_w^E=0.40$.  The limited-commitment allocation therefore exposes the more risk-averse wife to twice her efficient share of the innovation.  Its aggregate risk loading is
\[
    \eta_w\theta_w^2+\eta_m\theta_m^2=2.00,
\]
compared with $\eta_w\eta_m/(\eta_w+\eta_m)=1.20$ under full commitment.

\subsection{Continuous fertility}

Table \ref{tab:continuous} verifies the analytical comparative statics.  Increasing $\sigma$ from zero to $0.40$ halves the agreed fertility level.  At the baseline $\sigma=0.20$, insuring half of the unexpected loss raises $n^*$ from $0.353$ to $0.419$; complete insurance restores the deterministic-risk value $0.449$.  The full-commitment benchmark is $n_E=1.083$, compared with $n^*=0.353$.

\begin{table}[htbp]
\centering
\caption{Continuous fertility under career risk and policy}
\label{tab:continuous}
\begin{tabular}{cc@{\qquad}cc@{\qquad}cc}
\toprule
$\sigma$ & $n^*$ & Insurance $q$ & $n^*$ & Child subsidy $s$ & $n^*$\\
\midrule
0.00 & 0.449 & 0.00 & 0.353 & 0.000 & 0.353\\
0.10 & 0.419 & 0.25 & 0.388 & 0.010 & 0.370\\
0.20 & 0.353 & 0.50 & 0.419 & 0.025 & 0.397\\
0.30 & 0.284 & 0.75 & 0.441 & 0.050 & 0.443\\
0.40 & 0.225 & 1.00 & 0.449 & 0.100 & 0.541\\
\bottomrule
\end{tabular}
\begin{minipage}{0.91\textwidth}
\footnotesize\emph{Notes:} The risk column varies $\sigma$ with no insurance or subsidy.  The insurance and subsidy columns hold $\sigma=0.20$.  Policy units are normalized and are not fiscally comparable.
\end{minipage}
\end{table}

In the same baseline, increasing $\lambda$ from $0$ to $1$ lowers $n^*$ from $0.744$ to $0.179$.  This numerical monotonicity is not imposed: it arises because condition \eqref{eq:wife-margin} holds throughout the reported range.  Insuring half the innovation raises fertility by about $18.7$ percent at $\lambda=0.60$ and $22.6$ percent at $\lambda=1$.  This suggests a testable interaction: proportional insurance effects may be larger where post-birth resources track own earnings more closely.  The interaction is an illustrative numerical prediction, not an analytical theorem.

\subsection{Integer parity}

For a direct family-size interpretation, restrict $N\in\{0,1,2,3,4\}$.  The couple selects the feasible positive integer that maximizes the Nash product; it selects zero if no positive parity benefits both spouses.  We rescale child benefits to $b_m=1.50$ and $b_w=1.20$, leaving all other parameters unchanged.  Table \ref{tab:discrete} reports the results.

\begin{table}[htbp]
\centering
\caption{Discrete desired family size}
\label{tab:discrete}
\begin{tabular}{cc@{\qquad}cc@{\qquad}cc}
\toprule
$\sigma$ & $N^*$ & Insurance $q$ & $N^*$ & Child subsidy $s$ & $N^*$\\
\midrule
0.00 & 3 & 0.00 & 2 & 0.000 & 2\\
0.10 & 3 & 0.25 & 2 & 0.025 & 2\\
0.20 & 2 & 0.50 & 3 & 0.050 & 2\\
0.30 & 1 & 0.75 & 3 & 0.100 & 2\\
0.40 & 1 & 1.00 & 3 & 0.200 & 3\\
0.50 & 1 & & & &\\
\bottomrule
\end{tabular}
\begin{minipage}{0.91\textwidth}
\footnotesize\emph{Notes:} The policy columns hold $\sigma=0.20$.  Insurance and subsidy units are not fiscally comparable.  At the baseline, the full-commitment discrete choice is $N_E=3$, while the limited-commitment choice is $N^*=2$.
\end{minipage}
\end{table}

The discrete exercise produces economically visible thresholds without multiple continuous equilibria.  A moderate increase in career uncertainty can move desired family size from three to two, and a further increase from two to one.  At $\sigma=0.20$, coverage of half the innovation restores the three-child outcome.  A normalized child subsidy as large as $0.10$ does not cross that parity threshold in this example, whereas $s=0.20$ does.  Because program costs are unspecified, this is a mechanism comparison rather than evidence that one instrument is more cost-effective.

\section{Empirical discipline and quantitative extension}

The analytical core yields five empirical objects that can be measured or disciplined:

\begin{enumerate}
    \item the mean and variance of persistent earnings losses following birth, by pre-birth occupation, contract type, and cohort;
    \item the sensitivity of post-birth individual consumption or resource access to own earnings, which identifies $\lambda$ jointly with $\omega$;
    \item spouses' stated desired family size before and after the resolution of career uncertainty;
    \item heterogeneity in fertility responses to earnings-risk-reducing institutions, such as job-protected leave, re-entry guarantees, or earnings replacement;
    \item differential responses to predictable cash transfers and state-contingent career protection.
\end{enumerate}

A journal-ready quantitative model should replace the two-point normalized exercise with an estimated or externally calibrated distribution of the child penalty.  A tractable route is a finite-state career process $X\in\{x_1,\ldots,x_J\}$, integer parity, and backward induction over one pre-birth and one post-birth stage.  The analytical incidence formulas remain unchanged.  The quantitative section can then report policy elasticities, welfare, participation, and fiscal cost under a common budget.

The model has two natural identification predictions.  First, holding the mean penalty fixed, groups with more dispersed post-birth career outcomes should exhibit lower planned or realized parity.  Second, reducing dispersion should matter more when the wife bears a larger ex post share of her own earnings loss.  These are sharper than a generic prediction that higher expected female wages reduce fertility.

\appendix
\section{Proofs}

\subsection{Proof of Proposition 1}

For $i\in\{m,w\}$,
\[
    G_i'(n)=\frac{b_i}{1+n}-C_i-R_in,
    \qquad
    G_i''(n)=-\frac{b_i}{(1+n)^2}-R_i<0.
\]
If $b_i\leq C_i$, then $G_i'(0)\leq0$.  Strict concavity and $G_i(0)=0$ imply $G_i(n)<0$ for all $n>0$, so no positive agreement satisfies both participation constraints.

Now suppose $b_i>C_i$ for both spouses.  Each $G_i$ is positive in a right neighborhood of zero and, by strict concavity and the linear or quadratic cost term, has a unique positive root above which it is negative.  The common positive-surplus set is therefore a nonempty interval.  On its interior, the log Nash objective
\[
    \Phi(n)=\rho\log G_m(n)+(1-\rho)\log G_w(n)
\]
is strictly concave because
\[
    \Phi''(n)=\rho\frac{G_m''G_m-(G_m')^2}{G_m^2}
    +(1-\rho)\frac{G_w''G_w-(G_w')^2}{G_w^2}<0.
\]
Moreover, $\Phi(n)\to-\infty$ as $n$ approaches zero and as it approaches the upper boundary of the common positive-surplus set.  Hence there is a unique interior maximizer satisfying \eqref{eq:foc}. \hfill$\square$

\subsection{Proof of Proposition 2}

Let $H_i=G_i'/G_i$.  At the unique interior solution, $F_n=\Phi''(n)<0$.  Treating $C_i$ and $R_i$ as primitive coefficients,
\begin{align}
    \frac{\partial H_i}{\partial C_i}
    &=\frac{-G_i+nG_i'}{G_i^2}<0,\label{eq:HC}\\
    \frac{\partial H_i}{\partial R_i}
    &=\frac{n\left(-G_i+\tfrac12nG_i'\right)}{G_i^2}<0.\label{eq:HR}
\end{align}
To sign these expressions, strict concavity and $G_i(0)=0$ imply $G_i(n)>nG_i'(n)$.  This proves \eqref{eq:HC}.  If $G_i'\geq0$, the same inequality implies $G_i>\tfrac12nG_i'$.  If $G_i'<0$, that inequality is immediate because $G_i>0$.  Thus \eqref{eq:HR} also holds.

Since $\partial C_i/\partial\mu=\theta_i>0$,
\[
    F_\mu=\rho H_{m,C}\theta_m+(1-\rho)H_{w,C}\theta_w<0.
\]
The implicit-function theorem gives $n_\mu^*=-F_\mu/F_n<0$.  Likewise, with $v=\sigma^2$,
\[
    \frac{\partial R_i}{\partial v}=\eta_i\theta_i^2>0
\]
implies $F_v<0$ and $n_v^*=-F_v/F_n<0$. \hfill$\square$

\subsection{Proof of Proposition 3}

From \eqref{eq:thetaw}--\eqref{eq:thetam},
\[
    \theta_{w,\lambda}=1-\omega,
    \qquad
    \theta_{m,\lambda}=-(1-\omega).
\]
Therefore
\begin{align*}
    C_{w,\lambda}&=(1-\omega)\mu,
    &R_{w,\lambda}&=2\eta_w\theta_w(1-\omega)\sigma^2,\\
    C_{m,\lambda}&=-(1-\omega)\mu,
    &R_{m,\lambda}&=-2\eta_m\theta_m(1-\omega)\sigma^2.
\end{align*}
Using $A_i=-H_{i,C}>0$ and $B_i=-H_{i,R}>0$,
\[
    F_\lambda=(1-\omega)\left[\rho D_m-(1-\rho)D_w\right].
\]
Because $n_\lambda^*=-F_\lambda/F_n$ and $F_n<0$, $n_\lambda^*$ has the same sign as $F_\lambda$.  This proves \eqref{eq:lambda-sign} and \eqref{eq:wife-margin}. \hfill$\square$

\subsection{Proof of Proposition 4}

Let $v(q)=(1-q)^2\sigma^2$.  Proposition 2 gives $n_v^*<0$, while
\[
    v_q=-2(1-q)\sigma^2<0
\]
for $q<1$.  Hence $n_q^*=n_v^*v_q>0$.

For the child subsidy,
\[
    C_{w,s}=-\omega<0,
    \qquad
    C_{m,s}=-(1-\omega)<0.
\]
Since $H_{i,C}<0$, it follows that $F_s>0$.  Thus $n_s^*=-F_s/F_n>0$. \hfill$\square$

\section{Reproducibility note}

The numerical values in Tables \ref{tab:continuous} and \ref{tab:discrete} are generated by a standard-library Python implementation.  The continuous solution brackets the unique root of \eqref{eq:foc}; the discrete solution enumerates feasible parities.  Local finite-difference checks verify the signs in Propositions 2 and 4 at the baseline.

\end{document}